\documentclass[aps,pre,twocolumn,showpacs,superscriptaddress,10pt]{revtex4-1}
\pdfoutput=1
\usepackage{epsfig,graphicx,amsmath,amssymb}
\usepackage{natbib}
\usepackage[nooneline]{subfigure}
\usepackage{color}

\begin{document}
\title{Scale invariant Green-Kubo relation for time averaged diffusivity}

\author{Philipp Meyer}
\affiliation{
Max Planck Institute for the Physics of Complex Systems
Noethnitzer Str. 38
D 01187 Dresden
Germany}
\author{Eli Barkai}
\affiliation{Department of Physics, 
Institute of Nanotechnology and Advanced Materials, Bar-Ilan University, 
Ramat-Gan, 52900, Israel}
\author{Holger Kantz}
\affiliation{
Max Planck Institute for the Physics of Complex Systems
Noethnitzer Str. 38
D 01187 Dresden
Germany}

\date{\today}

\begin{abstract}

In recent years it was shown both theoretically and experimentally that 
in certain systems exhibiting
anomalous diffusion the time and ensemble average mean squared
displacement  are remarkably different. 
The ensemble average diffusivity is obtained from a
scaling Green-Kubo relation, which connects
the scale invariant non-stationary velocity correlation function 
with the transport coefficient.
Here we obtain the relation between time averaged diffusivity, 
usually recorded in single particle
tracking experiments, and the underlying scale invariant 
velocity correlation function.
The time averaged mean squared displacement is given by 
$\overline{\delta^2} \sim 2 D_\nu t^{\beta}\Delta^{\nu-\beta}$  
where $t$ is the total measurement time and $\Delta$ the lag time. 
Here $\nu>1$ is the anomalous diffusion exponent obtained from ensemble
averaged measurements $\langle x^2 \rangle \sim t^\nu$  
while $\beta\ge -1$ marks the growth or decline  of the kinetic 
energy $\langle v^2 \rangle \sim t^\beta$.
Thus we establish a connection between exponents which can be read off
the asymptotic properties of the velocity
correlation function and similarly for 
the transport constant $D_\nu$. 
We demonstrate our results with non-stationary scale invariant
stochastic and deterministic models, thereby highlighting that
systems with equivalent behavior in the ensemble average can differ 
strongly in their time average. 
This is the case, for example, if averaged kinetic energy is finite,
i.e. $\beta=0$, 
where $\langle \overline{\delta^2}\rangle \neq \langle x^2\rangle$.

\end{abstract}

\maketitle
\section{Introduction}

A central result of nonequilibrium statistical
physics is the Green-Kubo formalism.
It relates the diffusion constant $D$ of a normal diffusive system
to the stationary velocity correlation function 
$\langle v(t+\tau)v(t)\rangle$ of the process. 
The brackets $\langle ...\rangle$ denote the ensemble average.
The Green-Kubo relation reads \cite{Kubo}
\begin{equation}
D=\int_0^\infty d\tau \langle v(t+\tau)v(t)\rangle ,
\end{equation} 
where $v=dx/dt$.
In theoretical physics we mostly consider ensemble averages 
while in the real world it is sometimes
not possible to measure an ensemble because only one 
realization of a process is recorded.
In such systems we operate with the time average. 
In ergodic systems the time average mean squared displacement (TA MSD) 
$\overline{\delta^2}$, defined below, is the same as the ensemble average
(EA MSD) $\langle x^2(t)\rangle =2Dt$. 
So there is a unique way of defining the transport coefficient $D$
in the sense that the two procedures are equivalent. 

It is known that in complex and disordered systems
the TA MSD might depend on the total measurement time $t$ 
as well as on the lag time $\Delta$ \cite{PCCPralf}. 
Here we want to focus on non-stationary processes, 
with scale invariant correlation functions.
There are many examples for such non-stationary scale invariant processes
including the velocity of laser-cooled atoms \cite{Kessler}, 
the motion of a tracer particle in a
crowded environment \cite{Leibovich}, 
elastic models of fluctuating interfaces \cite{Taloni},
diffusion in heterogeneous environment \cite{Cherstvy} 
and blinking quantum dots \cite{Margolin}.

Since for scale invariant non stationary velocity correlation functions the 
EA MSD  $\langle x^2(t)\rangle =2D_\nu t^\nu$ is not equivalent to the 
TA MSD, we need two scaling Green-Kubo relations.
The one for the EA MSD was investigated previously \cite{DechantPRX}. 
Here we focus on the time averaged MSD of certain anomalous processes.
We should remark, that the focus of the work of 
Dechant et al. \cite{DechantPRX}
was on super diffusive processes
while the approach in fact also works for subdiffusive processes when
certain conditions on the exponents are met (see details below).
A scaling Green-Kubo relation for time averages is obtained in chapter 3.
Unlike for the EA MSD that was calculated in \cite{DechantPRX}, 
it is now for the TA MSD important to know the scaling of the EA 
$\langle v^2(t)\rangle$ of the underlying velocity. 
We assume no net drift and $\langle x(t)\rangle = \langle v(t) \rangle =0$.

This text is also a story about different models - stochastic and 
deterministic - that describe these processes. 
We will see differences in the results for such models which on the 
first sight look very similar.
The calculation of the EA MSD and TA MSD for all kinds of cases
and models has been of interest for scientists for a long 
time \cite{PCCPralf,r4,Phase,AlbersDis,LWRMP,Dani1,AkimotoSaito,HeBurov}.
In chapter 4 we apply the scaling Green-Kubo relation to a velocity 
renewal process and explicitly calculate the TA MSD. 
This was previously calculated by Tony Albers \cite{AlbersDis} for a 
similar model using a different technique.

In chapter 5 we compare our results to a jump model
which is a random walk description of anomalous transport.
It exhibits the same behavior in the EA MSD as the
renewal velocity process. 
However, the TA MSD in the two models 
are very different. This implies that the TA MSD is an observable 
sensitive to underlying paths, if compared with the EA MSD.

\begin{figure}
\includegraphics[width=0.5\textwidth]{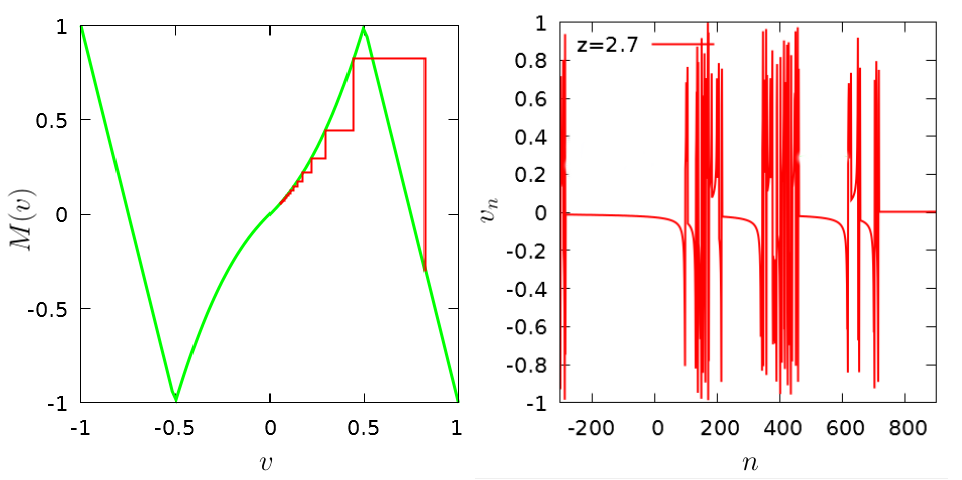}
\caption{The symmetrized Pomeau-Manneville map for $z=2.3$. 
A velocity trajectory is shown in the right panel.}
\label{fig01}
\end{figure}

In the end of this text we also want to introduce a third process which 
is completely deterministic. The idea of generating noise by deterministic 
chaotic processes goes back to \cite{Beck}.
One of the most prominent generators of anomalous dynamics in 
deterministic systems is the Pomeau-Manneville map (PM) \cite{Pomeau}. 
It shows intermittent behavior i.e. alternating between chaotic bursts 
and long waiting times with much slower dynamics. 
It was connected to aging \cite{Barkai03}, weak ergodicity breaking 
\cite{Bel} and anomalous diffusion \cite{r3}. 
We will be looking at pseudo-Brownian motion which is generated by a
modified version of the map.
The coordinate at discrete time $t$ is given by $x_t =\sum_{i=0} ^t v_i$.
Here $v_i$ is generated deterministically  with PM type of map
\begin{equation}
\label{eqPM01}
\begin{array}{l}
v_{n+1}= M(v_n) \\
M(v) =\left\lbrace\begin{array}{l l}
-4v-3 &\mbox{  for  } v<-1/2\\
v(1+|2v|^{z-1})&\mbox{  for  } |v|<1/2\\
-4v+3 &\mbox{  for  } 1/2<v
\end{array}\right. .
\end{array}
\end{equation}
Fig. \ref{fig01} shows the map and a typical trajectory.
The velocity in this model is bounded and  $-1<v<1$. Two 
back to back unstable  fixed points in the vicinity of $v=0$ 
imply long power law distributed  sticking times,
with small velocity, interrupted by bursts and then an injection 
back to the vicinity of the indifferent fix point,
i.e. this is the well known phenomenon of intermittency.
We want to compare this deterministic process to two different 
stochastic models which will be introduced in chapter IV and V.

\section{Scale Invariant Green-Kubo Relation}

We consider diffusive processes
with zero mean which in the long time limit
\begin{equation}
\label{eq00}
\langle x^2(t) \rangle = 2 D_\nu t^\nu
\end{equation}
and $\nu>0$. 
The coordinate of the particle is given by 
$x(t) = \int_0 ^t v(t') {\rm d} t'$ so initially $x(t)|_{t=0}=0$.
The random  velocity process $v(t)$ has zero mean.
The time averaged velocity  correlation function determined empirically
from  the velocity path recorded in the time interval $(0,t)$ is 
\begin{equation}
\label{eq02}
C_{{\rm TA}}(t,\tau) = {1 \over t-\tau} \int_0 ^{t-\tau} 
{\rm d} t_1 v(t_1) v(t_1 + \tau) . 
\end{equation}
In general this correlation function, which is a functional 
of the random process $v(t)$ is random,
namely specific to the underlying velocity  path. 
We assume that upon averaging the 
time averaged correlation function exhibits scale invariance namely
\begin{equation}
\label{eq03}
\langle C_{{\rm TA}} (t,\tau) \rangle = 
{\cal C}_{{\rm TA}} t^{\nu-2} \phi _{{\rm TA}} 
\left( {\tau \over t} \right) .
\end{equation}
The averaging is with respect to an ensemble of paths $v(t)$ which 
can contain an average over
initial conditions and stochastic histories, in experiment this 
correlation function is obtained from  an 
ensemble of measured  trajectories collected
under some specified  physical conditions. 
In physical systems such scale invariance is found in the scaling 
limit where both $\tau$ and $t$ are large. 
However for now we assume the scale invariance works for all
times which is an idealization. 
We now find the relation between the transport coefficient $D_\nu$ 
and the scale invariant time averaged correlation function.

First note that unlike stationary processes the time averaged 
correlation function is not identical to the ensemble average correlation
function even in the long time limit. The ensemble average correlation
function $C_{{\rm EA}}(t+\tau,t) = \langle v(t) v(t+\tau) \rangle$
also exhibits scale invariance
\begin{equation}
\label{eq04}
C_{{\rm EA}} (t+\tau,t) = {\cal C}_{{\rm EA}} t^{\nu-2} \phi_{{\rm EA}} 
\left( {\tau \over t} \right) .
\end{equation}
The correlation functions are related to one another according to
\cite{NavaPRE}
\begin{equation}
\label{eq05}
{\cal C}_{{\rm TA}} \phi_{{\rm TA}} ({q}) = {{q}^{\nu-1} \over 1- {q}} 
\int_{{q}/(1-{q})} ^\infty {\rm d} y{ {\cal C}_{{\rm EA}} \phi_{{\rm EA}} (y) 
\over y^\nu}.
\end{equation}
This relation can be easily derived from the definition of the time averaged
correlation function Eq. (\ref{eq02}). 

Since we have two related correlation functions, both of them can be 
used  to find the transport coefficient  $D_\nu$. 
The relation between $D_\nu$ and the ensemble
averaged correlation function was presented previously \cite{DechantPRX}.

The MSD is 
\begin{equation}
\label{eq06}
\langle x^2(t) \rangle = 2  \langle \int_0 ^{t} {\rm d} t_1 v(t_1) 
\int_{t_1}^t {\rm d}t_2  v(t_2)   \rangle.
\end{equation}
Switching variables of integration 
$t_1 = t_1$ and $t_2 = t_1 +\tau'$
we get
\begin{equation}
\label{eq07}
\langle x^2(t) \rangle = 2 
\int_0 ^t {\rm d} \tau' \left( t - \tau' \right) 
{\int_0 ^{t-\tau'} {\rm d} t_1 \langle  v(t_1) v(t_1+ \tau') \rangle 
\over t - \tau'} .
\end{equation}
Using the definition of the time averaged correlation function 
and ${q}=\tau'/t$ we find
\begin{equation}
\label{eq08}
D_\nu=
{\cal C}_{{\rm TA}} 
 \int_0 ^1 {\rm d} {q} (1 -{q}) 
\phi_{{\rm TA}}({q}) .
\end{equation}
Using Eq. (\ref{eq05}) and integration by parts we retrieve
\cite{DechantPRX}
\begin{equation}
\label{eq09}
D_\nu = {{\cal C}_{{\rm EA}} \over \nu} 
\int_0 ^\infty {\rm d} {q} (1 + {q})^{-\nu} \phi_{{\rm EA}} ({q}). 
\end{equation}
This is called a scaling Green-Kubo relation, since it connects
between the aging correlation function and $D_\nu$.
While Eqs. (\ref{eq08},\ref{eq09})
are clearly identical the appearance of two types of correlation
functions implies that these tools  should be used
with some care. Theoreticians usually focus on the ensemble average
correlation function, and then Eq. (\ref{eq09}) is useful,
but from data one may in principle obtain the time average
scaling function $\phi_{{\rm TA}}({q})$
and then Eq. (\ref{eq08}) is worthy.

In our derivation we assumed that the integrals in Eqs. 
(\ref{eq08},\ref{eq09}) are finite. This implies some limitations
on the shape properties of correlation functions, which will
soon be specified. We also assume that $\phi_{{\rm EA}}({q})$
is positive valued, and all examples will focus on monotonically
decaying functions. 
More explicitly we are limited by
\cite{DechantPRX}
\begin{equation}
\label{eq10}
\begin{array}{l l l l }
\phi_{{\rm EA}} ({q}) < c_1 {q}^{ - \delta_1} &
\mbox{with} & 2-\nu\le \delta_1 <1 & {q} \rightarrow 0 \\
\phi_{{\rm EA}} ({q}) <  c_u {q}^{ - \delta_u} &
\mbox{with} & \delta_u > 1-\nu & {q} \rightarrow \infty. 
\end{array}
\end{equation}
where $c_1>0$ and $c_u>0$ are some constants. 
It is emphasized that these conditions are inequalities, namely we
do not demand power law behaviors in the limits of ${q}\rightarrow 0$ 
and ${q}\rightarrow \infty$. 

Here we used the exponents characterizing the ensemble average correlation 
function. One can use in principle exponents characterizing the
time averaged correlation function instead. These exponents
are related to one another, for example if 
the ensemble averaged correlation function behaves like 
$\phi_{{\rm EA}}(q) \sim {q}^{-\delta_1}$ for ${q} \rightarrow  0$ 
so does the time averaged
scaling  function. This can be easily verified using Eq. 
(\ref{eq05}). Similarly the coefficients ${\cal C}_{{\rm TA}}$ is 
proportional to ${\cal C}_{{\rm EA}}$.
From now on we will use the ensemble average correlation
function $\phi_{{\rm EA}}({q})$. 

In the processes we consider below the variance of velocity is 
asymptotically
\begin{equation}
\label{eq11}
\langle v^2(t) \rangle \sim a {\cal C}_{{\rm EA}}  t^\beta
\end{equation}
and
\begin{equation}
\label{eq11a}
-1\le \beta<\nu-1.
\end{equation}
Note that in \cite{DechantPRX} the case
$\beta>0$ was considered, however the conditions for the theory to hold
are not as limiting, and in fact the case $-1<\beta<0$ will be important
in our example. The case $\beta=0$ is of course natural in systems where
the average kinetic energy of the particle is a constant. 
Now continuity demands 
\begin{equation}
\label{eq11aa}
\phi_{{\rm EA}}({q}) \sim
c_1 {q}^{-\delta_1} 
\end{equation}
for small ${q}$  with
\begin{equation}
\label{eq12}
\delta_1 = 2 - \nu + \beta .
\end{equation}
This is a useful relation since it gives the small ${q}$ behavior
of the correlation function in terms of exponents $\beta$ and $\nu$
which are both measurable.  In what follows we use the exponents $\beta$ 
and $\nu$ to find the properties of the time averaged diffusion constant.

\section{Time Averaged Mean Squared Displacement}

The time averaged MSD is 
\begin{equation}
\label{eq14}
\overline{\delta^2} \equiv {1 \over t - \Delta} \int_0 ^{t-\Delta} 
\left[ x(t_0+\Delta) - x(t_0)\right]^2 {\rm d} t_0.
\end{equation}
Here $t$ is the measurement time, namely the stochastic
path $x(t')$ is recorded in the time interval $(0,t)$
and $\Delta\ll t$ is the lag time.
For Brownian motion $\overline{\delta^2} \sim 2 D \Delta$ 
so the time average procedure yields the diffusion constant 
recorded in ensemble measurement
$\langle x^2(t)\rangle = 2 D t$. 
For scale invariant processes under consideration in this manuscript 
the identity of time and ensemble averages is broken.
Further, the time average may remain a random variable 
even in the long time limit \cite{PCCPralf,AkimotoSaito,HeBurov}. 
We will not address the fluctuations of this widely observed quantifier 
of diffusion processes, instead we focus on the ensemble average 
$\langle \overline{\delta^2}\rangle$.

Using Eq. (\ref{eq14}) and $t\gg \Delta$ we get
\begin{equation}
\label{eq15}
\begin{array}{c}
\langle \overline{\delta^2} \rangle \simeq  \\
{1 \over t} \int_0 ^{K \Delta} 
\langle \left[ x(t_0+\Delta) - x(t_0)\right]^2\rangle  {\rm d} t_0
+\\
{1 \over t} \int_{K \Delta} ^t 
\langle \left[ x(t_0+\Delta) - x(t_0)\right]^2\rangle  {\rm d} t_0 .
\end{array}
\end{equation}
Here $K$ is some large number satisfying $\Delta\ll K \Delta\ll t$.
It is clear that in the limit of $t\rightarrow \infty$
only the second integral contributes and the first is negligible. 
Further in the second integral
we have to find the  MSD recorded between 
time $t_0$ and $t_0+\Delta$ under the condition that $t_0\gg\Delta$. 
We denote
\begin{equation}
\label{eq16}
\begin{array}{c}
\langle \Delta x^2(\Delta) \rangle_{t_0} \equiv
\langle \left[ x(t_0 + \Delta) - x(t_0)\right]^2 \rangle =  \\
2 {\cal C}_{{\rm EA}} 
\int_0 ^{\Delta} {\rm d} t_2 \int_0 ^{t_2} {\rm d} t_1 
\left( t_1 + t_0\right)^{\nu -2} \phi_{{\rm EA}} 
\left( { t_2 - t_1 \over t_1 + t_0} \right).
\end{array}
\end{equation}
Since $\Delta\ll t_0$ it is clear that only the small ${q}$ behavior
of the correlation function is contributing to the integral in this limit,
hence using Eqs. (\ref{eq11aa},\ref{eq12}) one finds \cite{DechantPRX} 
\begin{equation}
\label{eq17a}
\langle \Delta x^2 (\Delta) \rangle_{t_0} \sim 
2 { c_1 {\cal C}_{{\rm EA}} 
\over \left( \nu - \beta - 1 \right) \left( \nu-\beta\right) }
(t_0)^\beta \Delta^{\nu-\beta}. 
\end{equation}

We can now derive our main equation in this section
inserting Eq. (\ref{eq17a}) in 
Eq. (\ref{eq15}) and performing a simple integral
\begin{equation} 
\label{eq17}
\langle \overline{\delta^2} \rangle \sim { 2 c_1 {\cal C}_{{\rm EA}} \over
\left( \beta  + 1 \right) \left( \nu- \beta - 1 \right) 
\left( \nu - \beta\right)} t^\beta \Delta^{\nu-\beta}.
\end{equation}
To determine the exponents of the time average mean square displacement
indirectly, for example via a measurement, 
one needs to know
$\beta$ which is a measure of the increase or decrease
of kinetic energy of the particle, 
and $\nu$ which as mentioned can be determined from ensemble 
averaged measurements of the MSD. Of course one can turn this around:
with the time average exponents and $\beta$ one can get $\nu$ and hence 
the ensemble averaged exponent.   
Here we see that the time averaged MSD is very different
from the ensemble average. It depends on the total measurement time $t$
and the lag time $\Delta$.  When $\beta=0$ meaning that 
$\langle v^2 \rangle$ is a constant, 
as one finds in normal thermal systems, 
and when $\nu=1$ as found for normal
transport, the time average behaves normally as expected
$\langle \overline{\delta^2}\rangle\propto\Delta$.
This case corresponds to the standard Green-Kubo relation and 
Eq. (\ref{eq17}) does not hold.
If kinetic energy is not increasing, i.e. $\beta=0$,
the time averaged MSD $\overline{\delta^2} \propto \Delta^\nu$
so it exhibits the same time dependence 
as does the ensemble average. 
Notice that unlike the ensemble average
MSD, where $D_\nu$ depends on the details 
of the correlation function namely on $\phi_{{\rm EA}}({q})$ 
in the range $0<{q}<\infty$
the time averaged MSD is determined
by $c_1 {\cal C}_{{\rm EA}}$ namely by the behavior of this 
function close to ${q}\rightarrow 0$.

\section{Renewal velocity  process}

We consider a renewal process $v(t)$ which will be used to demonstrate
the general theory derived so far. At random times
$t_0,t_1,t_2,....$ the particle experiences `strong collisions' 
in such a way that the velocity $v(t)$ is totally randomized, i.e., 
the correlation between the  velocities  of the  particle before and 
after a collision event is zero. 
Between the collision events particles
move deterministically. Let $n$ be the random  number of collision events
in the time interval $(0,t)$ and the process starts at the origin of time
namely $t_0=0$. Here time $t_n<t$ is the time when last stochastic
modification of velocity took place. At time $t$ the velocity is
\begin{equation}
\label{eq18}
v(t) = v_{\gamma,n} \gamma (t - t_n)^{\gamma-1}
\end{equation}
and we consider the case $0<\gamma$ (the case $\gamma<0$ is of interest,
at least mathematically and it can yield sub-diffusion). 
The case $\gamma=1$ implies motion at 
constant velocity between collision events.
Here $v_{\gamma,n}$ is a random variable with zero mean and finite variance.
The process starts at time $t=0$ and the waiting times between the renewal
events are independent identically distributed 
random variables with a common probability density function
(PDF) $\psi(\tilde{\tau})$ and $\tilde{\tau}>0$. 
Similarly the coefficients $v_{\gamma,j}$ are mutually independently
identically distributed random variables taken of a distribution 
$f(v_\gamma)$, with zero mean and finite variance
denoted $\langle (v_\gamma)^2\rangle$. 
So to describe the process we generate the pair 
$(\tilde{\tau}_0,v_{\gamma,0})$
say on a computer, and then for times shorter then $t_1=\tilde{\tau}_0$
the velocity is $v(t)= v_{\gamma,0} \gamma  t^{\gamma-1}$. 
Then the process is renewed, namely the pair 
$(\tilde{\tau}_1, v_{\gamma,1})$ is used and in the time interval 
$t_1<t<t_2 = \tilde{\tau}_0 + \tilde{\tau}_1$ the velocity is 
$v(t) = v_{\gamma,1}\gamma  (t - t_1)^{\gamma-1}$, etc.  
A trajectory of the process is presented in Fig. \ref{fig02}.

\begin{widetext}
\subsection{Velocity correlation function}

We now  investigate the velocity correlation function. 
We will focus on widely used fat tailed waiting time
PDFs
\begin{equation}
\label{eq20}
\psi(\tilde{\tau}) \sim A \tilde{\tau}^{-(1 + \alpha)} \ \ \mbox{when} \
\ \tilde{\tau} \rightarrow \infty
\end{equation}
and $\alpha>0$ for normalizability.
We note that even   
exponential statistics gives  certain strong  anomalies  when
$\gamma<0$, a case we do not study here.  
The mean waiting time is infinite if $0<\alpha<1$ while the
second moment of the same variable diverges when $0<\alpha<2$ 
(we will not consider marginal cases like $\alpha=1$ as they bring 
with them logarithmic corrections). 

\begin{figure}[ht]
\includegraphics[width=0.5\textwidth]{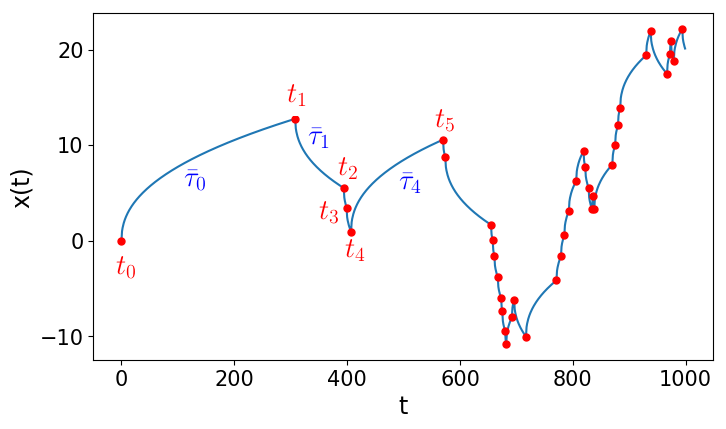}
\caption{A schematic diagram of a trajectory $x(t)$
for a renewal velocity process with 
$v_{\gamma,n}=\pm 1$ with equal probability. 
The parameters are $\alpha=5/14$, $\gamma=2/7$. 
The renewals occur at the times $t_0,t_1,\dots$ marked with red dots. 
The waiting times in between are denoted with 
$\bar{\tau}_0,\bar{\tau}_1,\dots$ and 
$t$ denotes the total measurement time.}
\label{fig02}
\end{figure}
We use methods similar to those used by Godreche and Luck 
\cite{GodrecheLuck}. Let 
$C_n(t,\tau)\equiv \langle v(t) v(t+ \tau)\rangle_n$ be the velocity 
correlation function  for a process with 
$n$ collisions in the time interval $(0,t)$.
We will obtain this function
and summation over $n$ will yield the sought after
correlation function $\langle v(t) v(t + \tau) \rangle = C(t,\tau)
=\sum_{n=0} ^\infty  C_n(t,\tau)$. 

 We define the double Laplace transform
\begin{equation}
\label{eq21}
C_n (s,u) = \int_0 ^\infty e^{ - s t} {\rm d} t 
\int_0 ^\infty e^{ - u \tau} {\rm d} \tau 
\langle v(t) v(t+ \tau)\rangle_n .
\end{equation}
The velocity is correlated only if $t_n < t < t+\tau<t_{n+1}$ 
namely when the observation times fall within the same epoch of travel. 
This is clearly the case since the velocities 
$v_{\gamma,j}$ are not correlated. 
We therefore have using Eq. (\ref{eq18})
\begin{equation}
\label{eq22}
\hat{C}_n (s,u) =
\langle v_{\gamma}^2 \rangle \gamma^2 \langle
\int_0 ^\infty e^{ - s t} {\rm d} t \int_0 ^\infty e^{ - u \tau} 
{\rm d} \tau (t-t_n)^{\gamma-1} (t + \tau -t_n)^{\gamma-1}
\theta\left( t_n < t < t_{n+1}\right) 
\theta(t_n < t_n + \tau<t_{n+1} )\rangle.
\end{equation}
Here $\theta(\cdots)=1$ if the condition in the parenthesis is valid otherwise 
the theta function  is zero, so $\theta(...)$ is a 
square pulse function. 
The average is with respect to the time process. 
Switching integration variables according to $t-t_n = y$ 
and using $t_{n+1} - t_n = \tilde{\tau}_n$ we find
\begin{equation}
\label{eq23}
\hat{C}_n (s,u) = \gamma^2 \langle (v_{\gamma})^2\rangle
\langle e^{ - t_n s} \int_0 ^{\tilde{\tau}_n} {\rm d} y\;  e^{ - s y} y^{\gamma-1}  
\int_0 ^{\tilde{\tau}_n - y} {\rm d} \tau\; e^{ - u \tau} \left( y + \tau \right)^{\gamma-1} \rangle.
\end{equation}
From renewal assumption the random variables $t_n$ and $\tau_{n+1}$ are
independent. Further since $t_n = \sum_{j=0} ^{n-1} \tilde{\tau}_j$ and
because the waiting times are also independent we have 
$\langle \exp( - s t_n)\rangle
= \hat{\psi}^n(s)$ where $\hat{\psi}(s)$ is the Laplace 
$\tilde{\tau} \rightarrow  s$ transform of $\psi(\tilde{\tau})$. 
The remaining average is with respect to 
$\tilde{\tau}_{n}$ which is a random variable drawn from 
$\psi(\tilde{\tau})$.
Hence we get
\begin{equation}
\label{eq24}
\hat{C}_n (s , u ) = \gamma^2 \langle (v_\gamma)^2 \rangle\hat{\psi}^n (s) 
\int_0 ^\infty {\rm d} \tilde{\tau } \psi(\tilde{\tau}) 
\int_0 ^{\tilde{\tau}} {\rm d} y e^{ - sy} y^{\gamma-1} 
\int_0^{\tilde{\tau}-y}
{\rm d} \tau e^{ - u \tau} \left( y + \tau\right)^{\gamma-1}.
\end{equation}
Let us denote  
\begin{equation}
\label{eq25}
W(\tilde{\tau} ) = \int_{\tilde{\tau}} ^\infty {\rm d} \tilde{\tau}'
\psi(\tilde{\tau}'), 
\end{equation}
which is the probability of not experiencing a renewal/collision 
in the time interval $(0, \tilde{\tau})$. Then  clearly
\begin{equation}
\label{eq26}
\hat{C}_n (s , u ) = 
\gamma^2 \langle (v_\gamma)^2 \rangle\hat{\psi}^n (s) 
\int_0 ^\infty {\rm d} \tilde{\tau } \left[-{{\rm d} \over {\rm d} 
\tilde{\tau}} W(\tilde{\tau})\right]
\int_0 ^{\tilde{\tau}} {\rm d} y e^{ - sy} y^{\gamma-1} 
\int_0 ^{\tilde{\tau}-y}
{\rm d} \tau e^{ - u \tau} \left( y + \tau\right)^{\gamma-1}.
\end{equation}
We now integrate by parts, and then the geometric series 
$\sum_{n=0} ^\infty \hat{\psi}^n(s)= 1/ [ 1 - \hat{\psi}(s)]$ gives
\begin{equation}
\label{eq27}
\hat{C}(s,u ) = 
{ \gamma^2 \langle (v_\gamma)^2 \rangle \over 1 -\hat{\psi}(s)}
\int_0 ^\infty {\rm d} \tilde{\tau} 
W(\tilde{\tau}) \tilde{\tau}^{\gamma-1} 
e^{ - u \tilde{\tau}} \int_0 ^{\tilde{\tau}} {\rm d}y 
e^{ - (s-u) y} y^{\gamma-1}.
\end{equation}
In principle the double inverse Laplace transform of this expression 
yields the velocity correlation function 
$\langle v(t) v(t+\tau)\rangle \equiv C(t,\tau)$.
We can invert from $u$ to $\tau$ rather easily, since 
the inverse Laplace transform of $\exp( - u x)$ 
is a delta function $\delta(\tau-x)$, hence we 
find
\begin{equation}
\label{eq28}
C(s,\tau) = {\gamma^2 \langle (v_\gamma)^2 \rangle \over 1 - 
\hat{\psi}(s) }
\int_{\tau} ^\infty {\rm d} \tilde{\tau} W(\tilde{\tau}) 
\tilde{\tau}^{\gamma-1 } \left(\tilde{\tau} - \tau\right)^{\gamma-1} 
e^{ - s( \tilde{\tau} - \tau) }.
\end{equation}
When $\gamma=1$ Eq. (\ref{eq27})
gives
\begin{equation}
\label{eq29a}
\hat{C}(s,u) = {\langle (v_1)^2 \rangle \over 1 - \hat{\psi}(s)}
 {  \hat{W}(s) - \hat{W}(u) \over u- s} ,
\end{equation}
where $\hat{W}(s)$ is the Laplace transform of $W(\tilde{\tau})$ and
from the convolution theorem $\hat{W}(s)  = [1 - \hat{\psi}(s)]/s$. 

\subsection{Long time limit with finite mean sojourn time}

We now classify behaviors of the correlation function in the limit of
long time $t$. We first consider the case when the average waiting
time $\langle \tilde{\tau} \rangle= \int_0 ^\infty 
{\rm d} \tilde{\tau} \tilde{\tau}  \psi(\tilde{\tau})$ is finite hence $\alpha>1$.
In this case the small $s$ behavior of $\hat{\psi}(s)$ is 
\begin{equation}
\label{eq29b}
\hat{\psi}(s) \sim 1 - \langle \tilde{\tau} \rangle s  
\end{equation}
and similarly for the small $u$ behavior of $\hat{\psi}(u)$.
The small $s$ behavior gives the large $t$ limit of the correlation function,
so using the small $s$ expansion
 $\exp[-s(\tilde{\tau} - \tau)] /[1 - \hat{\psi}(s)] \sim 
 1/(s \langle \tilde{\tau} \rangle )$ and inverting, i.e. the inverse Laplace transform
of $1/s$ is $1$, we find using 
Eq. (\ref{eq28})
\begin{equation}
\label{eq29c}
\lim_{t \to \infty} C(t,\tau) = { \gamma^2 \langle 
(v_\gamma)^2 \rangle \over \langle \tilde{\tau} \rangle} 
\int_{\tau} ^\infty {\rm d} \tilde{\tau} W(\tilde{\tau}) \tilde{\tau}^{\gamma -1} 
\left( \tilde{\tau} - \tau \right)^{\gamma -1} .
\end{equation}
When $\gamma=1$ 
\begin{equation}
\label{eq29cc}
\lim_{t \to \infty} C( t, \tau) = 
{\langle (v_1)^2 \rangle \over \langle \tilde{\tau} \rangle } 
\int_{\tau} ^\infty {\rm d} \tilde{\tau} W(\tilde{\tau}),
\end{equation}
which is known to the experts.

If the waiting time PDF is exponential with unit mean 
$\psi(\tilde{\tau})=\exp( -\tilde{\tau})$ for $\tilde{\tau}>0$ 
we have $\langle \tilde{\tau} \rangle =1$ and
\begin{equation}
\label{eq29d}
\lim_{t \to \infty} C(t,\tau) =  \gamma^2 \langle (v_\gamma)^2 \rangle 
\int_{\tau} ^\infty {\rm d} \tilde{\tau} 
e^{- \tilde{\tau}} \tilde{\tau}^{\gamma -1} 
\left( \tilde{\tau} - \tau \right)^{\gamma -1} .
\end{equation}
So when $\gamma<0$ the integral diverges, an indication to non-normal
behavior not investigated in this paper. For the case $\gamma=1/2$ and
$\langle (\gamma v_\gamma)^2 \rangle =1$ we get
\begin{equation}
\label{eq29e}
\lim_{t \to \infty} C(t,\tau) =   
e^{- \tau/2} K_0 (\tau/2). 
\end{equation}
This blows up at $\tau\rightarrow 0$ and in that limit
$\lim_{t \to \infty} C(t,\tau) \sim  - \log (\tau)$. So even
for exponential waiting times we get non trivial behaviors when 
$\gamma\le 1/2$ an effect which is related to the fact that for $\gamma<1$
the velocity blows up immediately after a renewal. 
The divergence of the correlation function at $\tau=0$ means
that the second moment of $v(t)$ defined in Eq. (\ref{eq18}) 
 is diverging, in reality this implies
that the variance is increasing with time $t$ and for any finite long
time we expect
a finite variance. 

Returning to fat tailed sojourn  time PDF Eq. (\ref{eq20})
we have in the limit of long waiting times
\begin{equation}
\label{eq29f}
W(\tilde{\tau}) \sim {A \tilde{\tau}^{-\alpha} \over \alpha}.
\end{equation}
Hence in the limit of long $\tau$ we get from Eq. 
(\ref{eq29c}) and $\alpha>1$ 
§
\begin{equation}
\label{eq29g}
\lim_{t \to \infty} C(t,\tau) \sim
 { \langle (\gamma v_\gamma)^2 \rangle \over \langle \tilde{\tau} \rangle} 
{ c_3 A \over \alpha} \tau^{2 \gamma - 1 - \alpha} .
\end{equation}
Here $2\gamma<1 + \alpha$ and
$c_3 = \int_1 ^\infty {\rm d} x 
x^{\gamma - 1- \alpha } (x - 1)^{\gamma-1}$. 
Eq. (\ref{eq29g}) is an example of a scale invariant correlation function,
which is of the non aging type.

So far we have considered the limit $t\to \infty$. As we have just shown,
this led to meaningless results in some cases, as the integrals diverge,
e.g. $c_3=\infty$ for $2 \gamma>1+\alpha$. 
So we consider the case when $t$ and also $\tau$ are long but finite. 
Returning back to Eq. (\ref{eq28}) we use the approximation, 
valid for large $t$ or small $s$,
\begin{equation}
\label{eq29h}
{ \exp\left[ - s \left( \tilde{\tau} - \tau\right)\right] \over 1 - 
\hat{\psi}(s) } \sim {\exp\left[ - s \left( \tilde{\tau} - \tau\right) 
\right] \over s \langle \tilde{\tau} \rangle} .
\end{equation}
Using convolution theorem of Laplace transform 
the inverse Laplace transform
$(s \rightarrow t)$ of the expression on the RHS is a pulse function
equal $1$ if $0<\tilde{\tau} - \tau< t$ otherwise it is zero.
Hence inverting Eq. (\ref{eq28}) in the limit of long $t$
\begin{equation}
\label{eq29i}
C(t,\tau) \simeq 
{ \langle ( \gamma v_\gamma)^2 \rangle \over \langle \tilde{\tau} \rangle} 
\int_{\tau} ^{t + \tau} {\rm d} \tilde{\tau} W(\tilde{\tau}) 
\tilde{\tau}^{\gamma-1} \left( \tilde{\tau} - \tau \right)^{\gamma-1} .
\end{equation}
When $\tau$ is large in such a way that $W(\tilde{\tau})$ 
for $\tilde{\tau} > \tau$, 
is described by 
Eq. (\ref{eq29f}) 
we find
\begin{equation}
\label{eq29j}
C(t,\tau) \simeq { \langle (\gamma v_\gamma)^2 \rangle 
\over \langle \tilde{\tau} \rangle } { A \over \alpha} 
\tau^{2 \gamma - 1 - \alpha} 
\int_1 ^{ 1 + t /\tau} {\rm d} x x^{\gamma - 1 - \alpha} 
(x-1)^{\gamma - 1}.
\end{equation}
This in the limit $t/\tau \rightarrow \infty$ recovers Eq. (\ref{eq29g})
when $2 \gamma<1 + \alpha$. 
If $t/\tau>>1$ and the conditions $2 \gamma> \alpha+1$, 
$\gamma>0$, $\alpha>1$ hold
$C(\tau,t) \propto t^{2 \gamma - 1 - \alpha}$.

Clearly Eq. (\ref{eq29j})
belongs to the class of scaling correlation functions described
by Eq. (\ref{eq04}).
We can now summarize and find the exponents and pre-factors
describing both the ensemble and time averaged transport. 
We have 
\begin{equation}
\label{eq29k}
{\cal C}_{{\rm EA}} = { \langle \left( \gamma v_\gamma\right)^2 \rangle 
A \over \alpha \langle \tilde{\tau} \rangle} 
\end{equation}
the ensemble averaged transport exponent is 
\begin{equation}
\label{eq29l}
\nu = 2 \gamma + 1 - \alpha>1.
\end{equation}
The ensemble averaged scaling correlation function
\begin{equation}
\label{eq29m}
\phi_{{\rm EA}} ({q}) = {q}^{2 \gamma - 1 - \alpha} \int_1 ^{1 + 1/{q}} 
{\rm d} x x^{\gamma-1- \alpha} (x-1)^{\gamma-1}.
\end{equation} 
In the small ${q}$ limit we have 
$\phi_{{\rm EA}}({q}) \sim c_1 {q}^{- \delta_1}$
as in Eq. (\ref{eq11aa}) and the exponent $\beta$ is obtained from
Eq. (\ref{eq12}). 
We find
\begin{equation}
\label{eqexpo01}
\begin{array}{l l l l}
\delta_1= - 2 \gamma + 1 +\alpha, \ &
\ \beta= 0, \ &  \ c_1=c_3, \ &  \ 
\mbox{if} \  \alpha<2 \gamma <1 + \alpha \\
\delta_1= 0, \ &  \ \beta= \nu-2=  2 \gamma -\alpha-1, \ &  \
c_1= ( 2 \gamma - \alpha-1)^{-1}, \ &   
\ \mbox{if} \  1+\alpha< 2\gamma . 
\end{array}
\end{equation}
The condition $\delta_u > 1- \nu$ also holds.
Note that the transition between the two behaviors is found for 
$2 \gamma=1 + \alpha$ and then $\nu=2$, so a qualitative transition takes
place when the ensemble averaged mean square displacement is ballistic. 
With  the information in Eq.
(\ref{eqexpo01})
 we may apply the scaling Green-Kubo relation
and predict the time averaged MSD.

\subsection{Long time limit with diverging  mean sojourn time}

We now consider the case $0<\alpha<1$ so here the mean flight time 
is infinite. 
The small $s$ expansion of the Laplace transform of the 
PDF of waiting times is
\begin{equation}
\label{eq29o}
\hat{\psi}(s) \sim 
1 - { A \Gamma(1 - \alpha) \over \alpha} s^\alpha + \cdots
\end{equation}
where $\Gamma(.)$ is the Gamma function. 
In the limit of large $\tau$ we insert Eqs.
(\ref{eq29f},\ref{eq29o}) in 
Eq. (\ref{eq28}) 
\begin{equation}
\label{eq29p}
C(s,\tau) \simeq { \langle \left( \gamma v_\gamma\right)^2\rangle 
\over \Gamma(1 - \alpha)} \int_{\tau} ^\infty {\rm d} \tilde{\tau} 
\tilde{\tau}^{\gamma - 1- \alpha} 
{ e^{ - s(\tilde{\tau} - \tau)} \over s^\alpha } 
\left( \tilde{\tau} - \tau \right)^{\gamma - 1},
\end{equation}
an asymptotic equation that does not depend on the amplitude $A$ or 
any other detail on $\psi(\tilde{\tau})$ besides the exponent $\alpha$.
To invert this formula 
from $s$ to $t$ we use the convolution theorem of Laplace transform,
the Laplace pairs 
$1/s^\alpha \leftrightarrow t^{\alpha-1} /\Gamma(\alpha)$,
$\exp[ - s(\tilde{\tau} - \tau)]\leftrightarrow \delta[ t - 
(\tilde{\tau} - \tau) ]$ to  find
\begin{equation}
\label{eq29q}
{ e^{ - s(\tilde{\tau} - \tau) } \over s^\alpha} \leftrightarrow 
\left\lbrace
\begin{array}{l l}
0 & \tilde{\tau} - \tau < 0 \\
{1 \over \Gamma(\alpha)[ t - (\tilde{\tau} - \tau)]^{1-\alpha}} 
& \ \mbox{otherwise} \\
0 & t<\tilde{\tau} - \tau.
\end{array}
\right.
\end{equation}
\begin{figure}[ht]
\begin{minipage}{0.48\textwidth}
\includegraphics[width=\textwidth]{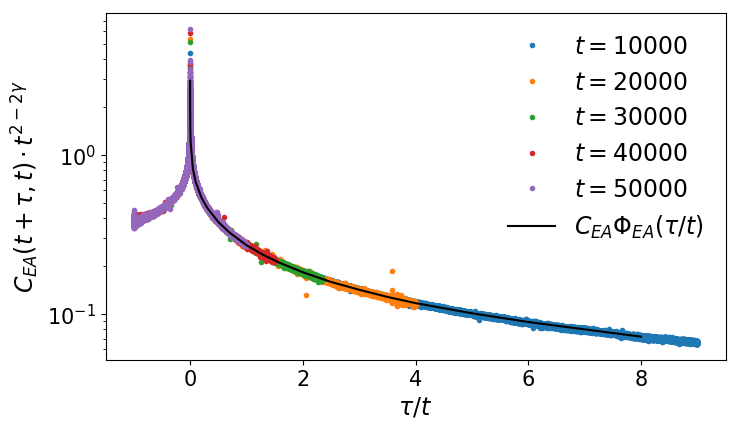}
\end{minipage}
\begin{minipage}{0.48\textwidth}
\includegraphics[width=\textwidth]{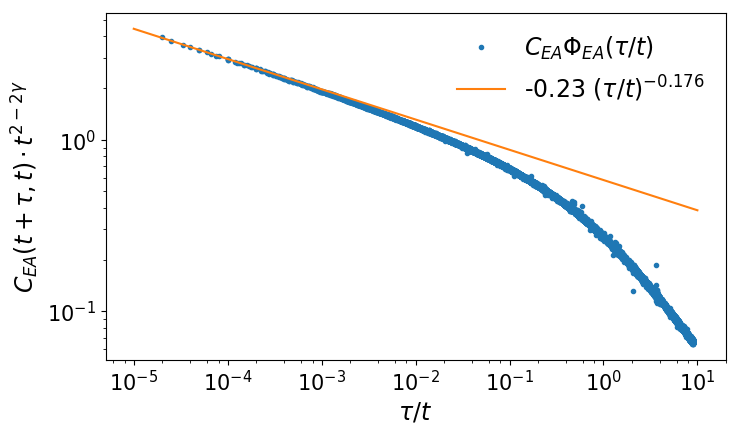}
\end{minipage}
\caption{Scaled correlation function of a renewal velocity process with 
$v_{\gamma,n}=\pm 1$ for $\alpha=5/13$, $\gamma =8/13$. 
The abscissa shows the fraction ${q}=\tau/t$ while the ordinate is rescaled 
by $t^{2-\nu}$. Thus the graph shows the function $\Phi_{\rm EA}$ 
(Eq. (\ref{eq29r1})), multiplied by $\cal{C}_{\rm EA}$. 
Simulations for different times nicely match to theory.
Left: the theory curve is drawn in black. It is universal for different $t$. Right:
log-log plot shows behavior for $\tau /t\rightarrow 0$. The fitted scaling 
exponent corresponds to the parameter $\delta_1$ and is therefore
directly related to $\beta$. It matches reasonably well with
the theory from Eq. (\ref{eq29t}) which predicts $\delta_1 =0.154$.}
\label{fig03}
\end{figure}
This is used to invert Eq. 
(\ref{eq29p})
\begin{equation}
\label{eq29r}
C(t,\tau) \simeq {\langle  \left( \gamma v_\gamma\right)^2 \rangle \over
\Gamma(1 - \alpha) \Gamma(\alpha) } 
t^{ 2 \gamma - 2} 
\phi_{{\rm EA}} ({q})
\end{equation} 
with 
\begin{equation}
\label{eq29r1}
\phi_{{\rm EA}} ({q}) = 
\int_{{q}} ^{1+ {q}} {\rm d} x x^{\gamma -1-\alpha}  
[1- (x-{q})]^{\alpha-1} (x-{q})^{\gamma-1}
\end{equation}
and ${q}=\tau/t$.
It is now easy to read off this equation  the expressions relevant
for the calculation of the time averaged  mean square displacement.
In particular 
using Eq. (\ref{eq04})
the anomalous diffusion scaling exponent
is 
\begin{equation}
\label{eq29s}
1<\nu=2 \gamma
\end{equation}
and clearly  ${\cal{C}_{{\rm EA}}} = \langle (\gamma v_\gamma )^2\rangle 
/ [ \Gamma(1-\alpha) \Gamma(\alpha)] = 
\langle (\gamma v_\gamma )^2\rangle \sin(\pi\alpha)/\pi$.
We then find
\begin{equation}
\label{eq29t}
\begin{array}{ l l l  l}
\delta_1= 0, \  & \beta=\nu-2=2\gamma-2, \ &   c_1 = 
\int_1 ^\infty {\rm d} y y^{ - 2 \gamma + 1} (y-1)^{\alpha-1}, 
& \ \mbox{if} \ 2 \gamma> 1 + \alpha \\
\delta_1=\alpha+1 - 2 \gamma, \ &  \beta= \alpha-1, \ & c_1 =
[1 + \alpha - 2 \gamma]^{-1} , \ & \ \mbox{if} \ 2 \gamma<1 +\alpha.  
\end{array}
\end{equation}
A plot of the scaled correlation function is shown in Fig. \ref{fig03}.
Again with this information the scaling Green-Kubo formalism
predicts the behaviors of the time averaged MSD, of course under the
conditions that the theorem holds, e.g. $-1<\beta$ and $\nu-\beta>1$.
For large ${q}$,
$\phi_{{\rm EN}} \propto {q}^{-\delta_u}$ with 
$\delta_u = 1 + \alpha-\gamma$ so
the  condition $\delta_u > 1 - \nu$ in Eq. (\ref{eq10}) holds.  

For ballistic Levy walks we have $\gamma=1$ and then the
condition $2 \gamma> 1 + \alpha$ holds (since here  $\alpha<1$)
and hence in that case we have only one type of behavior 
[the first line in Eq. (\ref{eq29t})]. So for Levy walks  $\nu=2$,
$\beta=0$, $c_1= \pi csc (\pi \alpha )$ and using a  well known identity
for Gamma functions  ${\cal C}_{{\rm EA}} c_1= \langle (v_1)^2 \rangle$.
Hence we find using Eq. (\ref{eq17})
$\langle \overline{\delta^2} \rangle \sim \langle (v_1)^2 \rangle
\Delta^2$. This was obtained in \cite{Dani0,Dani1}
and there also the corrections to this formula were investigated, 
as well as the fluctuations of the time average $\overline{\delta^2}$.

\subsection{Phase Diagram}

Now with the information on the exponents $\beta$ describing variance of
velocity and $\nu$ describing the variance of position in ensemble 
averaged sense, we  easily obtain the phase diagram of the 
time averaged mean square displacement using Eq. (\ref{eq17}).
We focus on $0<\gamma<2$ then in the case of diverging averaged waiting
time $0<\alpha<1$ and  using Eq. (\ref{eq29s},\ref{eq29t}) we find
\begin{equation}
\label{eq29u}
\langle \overline{\delta^2} \rangle \propto \left\lbrace
\begin{array}{l l}
t^{\alpha-1} \Delta^{2 \gamma -\alpha  + 1} \  & \
\mbox{max} (0,2 \gamma-1)<\alpha<\mbox{min} ( 1,2 \gamma) \\
t^{2 \gamma - 2} \Delta^2 \  & \  0 <\alpha<\mbox{min} (1 ,2 \gamma-1) 
\end{array}
\right. .
\end{equation}
This was obtained in \cite{AlbersDis}, where a CTRW approach was used.
We see that the motion is super-ballistic or ballistic and the time
average may either increase (an effect called rejuvenation) 
or decrease (called aging)  with total measurement time.
As mentioned it is controlled by the value of the exponent $\beta$,
describing time dependence of the kinetic energy.
For the case of finite average sojourn time $1<\alpha<2$ but diverging 
variance we use Eqs. (\ref{eq29l},\ref{eqexpo01}) to find
\begin{equation}
\label{eq29v}
\langle \overline{\delta^2} \rangle \propto \left\lbrace
\begin{array}{l l}
t^{0} \Delta^{2 \gamma + 1 - \alpha} \  &  \ \mbox{max}(1, 2 \gamma-1)< 
\alpha< \mbox{min} (2 , 2 \gamma) \\
t^{2 \gamma -\alpha -1} \Delta^2 \  &
\ 1 < \alpha<\mbox{min}(2 \gamma-1,2)
\end{array}
\right. .
\end{equation} 

\end{widetext}

\section{Random Walks}

The velocity  process under investigation
is related to the coupled  continuous time random walk model
\cite{Bouchaud,Review,Haus,Meerschaert,Dentz,Kutner,Zaburdaev}.
We now investigate a closely related random walk, 
which is not based on a velocity picture. 
Our goal is to show that the TA in both models are non-identical 
(unlike the ensemble averages). 
This implies that while random walk theory can work well in the 
ensemble average sense, when it comes to predictions of time averages 
it must be used with care.

\begin{figure}
\includegraphics[width=0.5\textwidth]{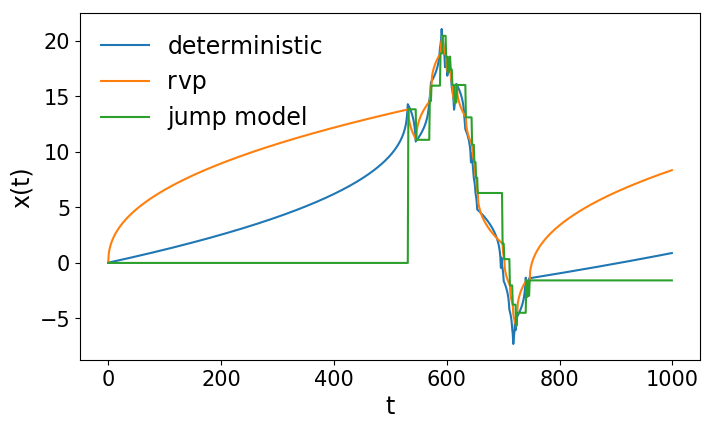}
\caption{Trajectories of all three processes: 
A deterministic process generated by the Pomeau-Manneville map, 
a renewal velocity process (rvp) with $v_{\gamma,n}=\pm 1$ and a random walk. 
The parameters are $z=2.8$, $\alpha=5/14$, $\gamma=2/7$. 
At the renewal times both stochastic processes are identical and the
deterministic model is also approximately the same. 
In between these points the trajectories behavior is very different. 
The deterministic trajectory lies in between the two others.}
\label{fig04}
\end{figure}

A random walker
waits localized in space and then makes a jump. The waiting times
in this well known model are independently identically distributed 
random variables drawn from the PDF $\psi(\tau)$. The size of each 
independent spatial  step is $\chi$ and we treat the case of equal 
probabilities of jumping to the left or right (no bias). 
In coupled processes the jump length and the waiting times are 
correlated and 
an example of their joint PDF is
\begin{equation}
\label{eq29}
\psi(\chi,\tau) = \psi(\tau) {1 \over 2} \left[ \delta(\chi-\tau^\gamma) 
+ \delta(\chi + \tau^\gamma) \right].
\end{equation}
This seems at first glance very similar to our process when the velocities 
$v_{\gamma,n}$ are
either $+1$ or $-1$ with equal probability. Indeed the random
size $\chi_j$ of displacement $j$  in the  velocity model is 
$\chi_j=v_{\gamma  j} (\tilde{\tau}_j)^\gamma$
so displacements in both approaches  are
identical (the displacement in the velocity model is simply the length 
traveled between renewal events). However the two models differ in the
path in between renewal events. In our case the velocity is always finite 
(unless $v_{\gamma,j}$ is zero). In the coupled CTRW particles wait 
and then jump, so the velocity is nearly always zero 
(see Fig. \ref{fig04}).
Still position of the particle at time $t_n$ is the same in both models. 
Hence the distribution of the position at this time is identical. 
Also the ensemble averaged MSD shows the same scaling as we see when 
comparing the results in \cite{Phase} to Eq. (\ref{eq29s})
\begin{equation}
\label{eqRW2}
\langle x^2\rangle \approx 
\left\lbrace\begin{array}{l l}
\frac{\Gamma(2\gamma-\alpha)}{|\Gamma(-\alpha)|\Gamma(1+2\gamma)}
t^{2\gamma} & \mbox{if }2\gamma>\alpha\\
\frac{\langle \tau^{2\gamma}\rangle}{c\Gamma(\alpha +1)} 
t^{\alpha} & \mbox{if }2\gamma<\alpha
\end{array}\right.
\end{equation}
for $0<\alpha <1$.

However, as we show below the time averaged MSD
in  these models are very different. In particular the exponents 
describing the time averaged
MSD in both models are not the same. 
We write the velocity of a jump model in the (not rigorous) form
\begin{equation}
v(t)=\sum_{\lbrace i:t_i<t\rbrace} \chi_i \delta(t-t_i)
\end{equation}
and assume that its ensemble average of the square scales like 
$\langle v^2(t)\rangle\propto t^\beta$.
We further assume that the scale invariant Green-Kubo relation 
holds for such processes.
Then we see that the TA MSD of such a jump process always depends 
linearly on $\Delta$.
The EA MSD
\begin{equation}
\langle x^2(t)\rangle = \left\langle\int_0^t dt_1 v(t_1) 
\int_0^t dt_2 v(t_2) \right\rangle
\end{equation}
can be rewritten with the $\theta$-step function a the integral over 
$\delta(t)$
\begin{equation}
\langle x^2(t)\rangle = \left\langle\sum_{ij} \chi_i \chi_j 
\theta(t-t_i) \theta(t-t_j)\right\rangle .
\end{equation}
Using $\langle \chi_i \chi_j\rangle =\delta_{ij}$, this yields
\begin{equation}
\langle x^2(t)\rangle =\left\langle \left( \int dt v(t) 
\right)^2 \right\rangle=\int dt \langle (v(t))^2 \rangle 
\propto t^{\beta +1}
\end{equation}
and therefore $\beta =\nu -1$. Using the ensemble averaged 
exponent $\nu$ given in Eq. (\ref{eqRW2}) together with
$\beta$ we find using Eq. (\ref{eq17})
\begin{equation}
\langle \overline{\delta^2}\rangle \sim
\left\lbrace\begin{array}{l l}
t^{2\gamma-1}\Delta & \mbox{if }2\gamma>\alpha\\
t^{\alpha -1}\Delta & \mbox{if }2\gamma<\alpha
\end{array}\right. .
\end{equation}
These equations were derived previously in \cite{Phase} from the 
underlying random walk. 
Here we have demonstrated that we can easily get $\beta$ and 
once $\nu$ is known predict the behavior of the TA MSD.

The result differs from the velocity model (Eq. \ref{eq29u}) not only in the
pre-factors but also in the exponents.
Thus the time average being sensitive to the whole shape of the path,
needs a precise definition of the model. 
And in this sense simplified random
walks which neglect the details of the velocity path can be 
widely different if compared to velocity models. 
The second difference between models is the 
propagation between last renewal event and measurement time $t$.
In wait and then jump process the particle is stuck in this last interval,
while in the velocity model it continues traveling. 
For long tailed PDF $\psi(\tilde{\tau})$ the statistics of this last 
traveling event is known to be of importance, 
for the calculation of quantities like $D_\nu$, however
this effect does not modify exponents describing the ensemble averaged
mean square displacement.

\section{The deterministic system}

We will now look at the deterministic process defined in (\ref{eqPM01}) 
and compare it to the two stochastic processes above.
\begin{figure}[ht]
\includegraphics[width=0.5\textwidth]{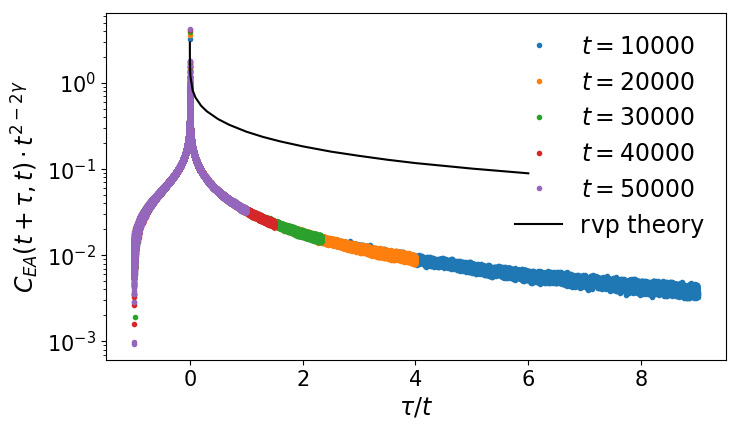}
\caption{Scaled correlation function of the deterministic process with 
$z=3.6$ ($\alpha=5/13$, $\gamma=8/13$). 
The abscissa shows the fraction $z=\tau/t$ while the ordinate is 
rescaled by $t^{2-\nu}$. 
The black curve is proportional to the function $\Phi_{EA}$ of 
the renewal velocity process (rvp), that was already shown in Fig. \ref{fig03}. 
It strongly differs from the numerical result of the deterministic process.}
\label{fig05}
\end{figure}
At time $t=0$ the initial velocity is uniformly  distributed in the 
interval $[-1,1]$. 
We generate on the computer the velocity path, just by iterating the map, 
and from this we obtain the coordinate $x_t$ 
and hence the time average MSD. 
The symmetry of initial conditions and of the map itself insures that
we do not have any drift.  
In Fig. \ref{fig05} we see that the correlation function of the 
map is scale invariant. 
We have chosen $z$ such that $\alpha=0.59$ and $\gamma=0.41$ 
(see next section) thus the scale invariant Green-Kubo relation holds.

\subsection{The connection between the parameter $z$, 
the waiting times and the jump lengths}
 
Let us first get an idea about waiting times and jump lengths in the
process. Roughly speaking, 
in the long time limit, properties of the map in the vicinity of the
unstable fixed point determine the statistical behavior of the paths. 
For $v>0$ the map is approximated with a differential 
equation
\begin{equation}
\label{eqPM02}
{{\rm d} v \over {\rm d} t} = \tilde{a} v^{z} \  \ \mbox{when} 
\ v \rightarrow 0,
\end{equation}
with $\tilde{a} > 0$ and for the specific example in 
Eq. (\ref{eqPM01}) $\tilde{a} = 2$.
Let $\alpha=1/(z-1)>0$, it is well known 
that the PDF of sojourn times in the 
vicinity of the indifferent points is given by \cite{Geisel}
\begin{equation}
\label{eqPM03}
\psi(\tilde{\tau}) \propto (\tilde{\tau})^{-1 - \alpha}.
\end{equation}
This is obtained from Eq. (\ref{eqPM02}). It is easy to integrate
Eq. (\ref{eqPM02}),
\begin{equation}
\label{eqPM02a}
 {\alpha \over \tilde{a}} \left[ 
{1 \over (v_0)^{1/\alpha} } - {1 \over (v_t)^{1/\alpha}} \right]=t
\end{equation}
and calculate the time it takes $v_t$ starting on $v_0$ to hit $v_b$. 
Here $v_b$ is some small constant on which   roughly speaking 
the continuous approximation of the map breaks down, say $v_b=1/4$.
Importantly it is an irrelevant parameter in the sense that 
eventually our results do not depend on its specific value. 
Using uniform distribution of initial conditions $v_0$,
one obtains Eq. (\ref{eqPM03}) which is well backed by simulations 
and theory.

 We now need to find the exponent $\gamma$. 
First note that the displacement during  
a renewal interval of length $\tilde{\tau}$, which we denote $\chi$ is 
by definition statistically  proportional to $\tilde{\tau}^{\gamma}$.
For the map we need to find 
\begin{equation}
\label{eqPM04}
\chi= \int_0 ^{\tilde{\tau}} v_t {\rm d} t.
\end{equation}
where $v_0$ is the injection point, marking the start of the escape 
from the unstable point, and $\tilde{\tau}$ is the time to reach the
boundary (when reinjection is taking place again). 
Using the fact that initial velocity $v_0$ 
is much smaller than its boundary value
$0<v_0<<v_b$ we find the scaling relation between the velocity at 
the start of the renewal (the injection point) and the time until 
particle hits the boundary
$v_0 \simeq [ \alpha/ (\tilde{a} \tilde{\tau})]^\alpha$.   
Of course not all the injection points $v_0$ are far from $v_b$,
however those injection events which land in vicinity of $v_b$
quickly escape and do not control the long time limit of the 
problem under investigation.
Here we use $v_t>0$, however in the dynamics
generated by the map both positive and negative velocities
are equally probable, and the sign of the velocity is determined merely 
by the injection point, namely does it happen to fall to the left or 
right of $v=0$. 
Inserting Eq. (\ref{eqPM02a}) 
in Eq. (\ref{eqPM04})  we find that in statistical sense 
\begin{equation}
\label{eqPM05}
\chi \propto { \alpha \over \tilde{a} (1 - \alpha)} \left( {\tilde{\tau} 
\tilde{a} \over \alpha} \right)^{1-\alpha}.  
\end{equation}
Hence to summarize we have that the non linear parameter of the map 
$z>1$ gives
\begin{equation}
\label{eqPM06}
\begin{array}{c c}
\alpha = {1 \over z -1}, \ \   \  & \ \ \   \gamma = { z-2  \over  z-1 }
\end{array} .
\end{equation}

\begin{figure}
\includegraphics[width=0.5\textwidth]{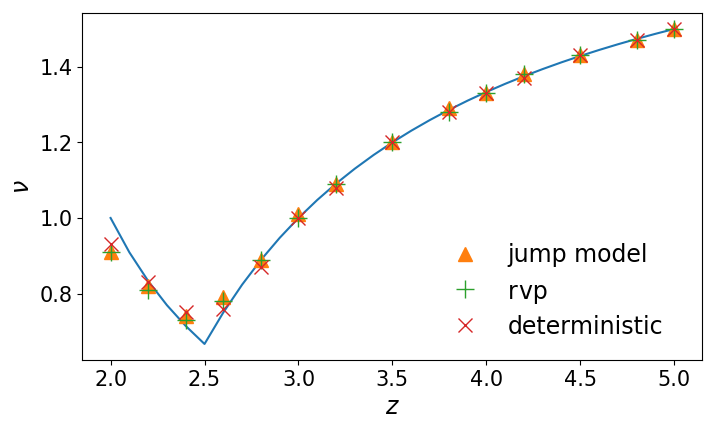}
\caption{The scaling exponent $\nu$
of the ensemble averaged mean squared 
displacement for all three processes.
The solid line illustrates the theory (Eq. (\ref{eqPM07}),\cite{Phase}).
The scaling exponent was fitted for times $10^4<t<10^6$. 
The averaging for all three models was done over 10000 trajectories.
}
\label{fig06}
\end{figure}
Using Eq. (\ref{eq29s}) we find
\begin{equation}
\label{eqPM07}
\nu = 2 \left( { z - 2 \over z-1} \right) 
\end{equation}
for the parameter range where our theory is applicable. 
As we will see this is true for $z>2.5$.
In Fig. \ref{fig05} the prediction Eq. (\ref{eqPM07}) 
is tested as we plot $t^{2-\nu} {\cal{C}}_{EA}(t + \tau, t)$ 
versus $\tau/t$ observing a data collapse. 
In Fig. \ref{fig06} we see the scaling exponents 
of the EA MSD of all three processes compared to each other. 
They are all the same. 
For $z<2$ ($\alpha>1$) the mean sojourn time is finite and therefore 
the processes exhibit normal diffusion. 
The range $2<z<2.5$ can also be understood. 
Here the statistics is completely dominated by the 
waiting times and the jump sizes/ velocities are small. 
It was investigated for example in \cite{Phase} 
and is also related to the spatial diffusion of the Pomeau-Manneville 
map described in \cite{Geisel}.

\subsection{The exponent $\beta$ and infinite ergodic theory}

\begin{figure}[ht]
\includegraphics[width=0.5\textwidth]{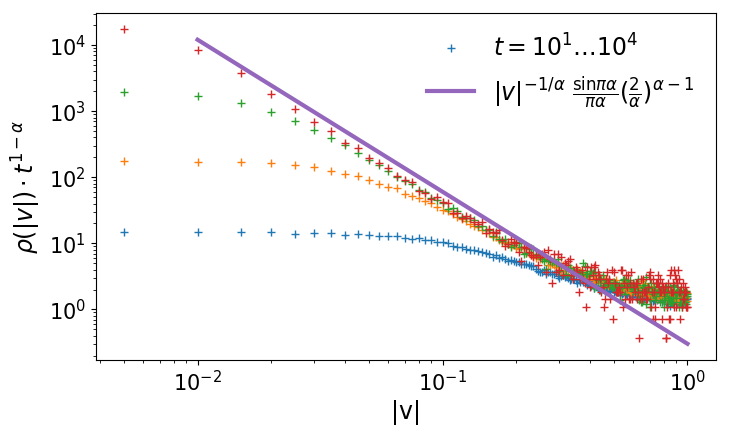}
\caption{Infinite invariant density of the Pomeau-Manneville map 
(also see in Korabel and Barkai \cite{Korabel1}) for 
$z=3.3$. As described in Eq. (\ref{eqINF01}) the rescaled density 
approaches Eq. (\ref{eqINF02}).
The averaging was done over 10000 trajectories.}
\label{fig07}
\end{figure}

The standard setting of the classical
Green-Kubo relation is for a system  in contact with a heat bath
and then the velocity distribution is Maxwelian. 
The processes under study are certainly
non-thermal and as we now demonstrate the velocity fluctuations
are described by infinite ergodic theory provided that 
$0<\alpha=1/(z-1)<1$.
Here we will use this theory to derive $\beta$ in a direct way.
Let $\rho(|v|,t)$ be the normalized
density of the variable $|v|$ at time $t$. 
This density is in principle obtained with the PM transformation 
starting from a smooth density
say a uniform density in $|v|<1$  (by smooth we mean in particular
that initially the density does not contain a delta function). 
For standard ergodic transformations and 
in the long time limit this density will converge to a normalisable 
invariant density, for example for the PM map when $\alpha>1$.
 However, when $\alpha<1$ the PM transformation
is analyzed with 
an infinite, i.e. non normalisable,  density $\rho_{{\rm inf}}(|v|)$ 
{(see Fig. \ref{fig07})}. 
The infinite density is related to the normalized density according to
\begin{equation}
\label{eqINF01}
\rho_{\rm \inf}(|v|) \sim t^{1-\alpha} \rho(|v|,t) .
\end{equation}
In fact now it becomes clear that the object on the left hand side 
is not normalisable, since the integral over $|v|$ of $\rho(|v|,t)$ 
is unity, and $t^{1-\alpha} \rightarrow \infty$ when $0<\alpha<1$. 
This definition can be used to estimate the infinite density 
from numerical data 
(mathematician usually define the infinite density up to a non 
specified constant, but here we follow convention in
\cite{Korabel1,Korabel2} ). It can be shown that 
\begin{equation}
\label{eqINF02}
\rho_{\rm inf} (|v|) = 
|v|^{- 1/\alpha} { \sin \pi \alpha \over \pi \alpha} 
\left( { \tilde{a} \over \alpha} \right)^{\alpha-1} h(|v|).
\end{equation}
Here $h(|v|)$ is a bounded function of order of unity, 
and most importantly $h(0)=1$. 
This means that the infinite density has a singularity close to 
$|v|\rightarrow 0$ namely  $\rho_{{\rm inf}} (|v|) \sim |v|^{-1/\alpha}$.
Consistently for  $0<\alpha<1$ the infinite 
density is clearly not integrable.
The beauty of infinite ergodic theory is that one may still
construct an ergodic theory based on this non-normalized function. 
Here we must distinguish between observables integrable and 
non-integrable with respect
to the infinite density. In our study we need the second moment of 
the velocity  found
using Eq. (\ref{eqINF01})
\begin{equation}
\label{eqINF03}
\langle |v|^2 \rangle = 
{ \int_0^1 |v|^2 \rho_{{\rm inf}}(|v|){\rm d} |v| \over t^{1 - \alpha}}.
\end{equation}
Using Eq. (\ref{eqINF02})
the integral is finite, meaning that the observable is integrable 
with respect to the infinite density if $\alpha>1/3$ or $z<4$. 
As long as $2<z<4$ we have from Eq. (\ref{eqINF04})
$\beta=\alpha -1= (2-z)/(z-1)$.

\begin{figure}[htb]
\includegraphics[width=0.5\textwidth]{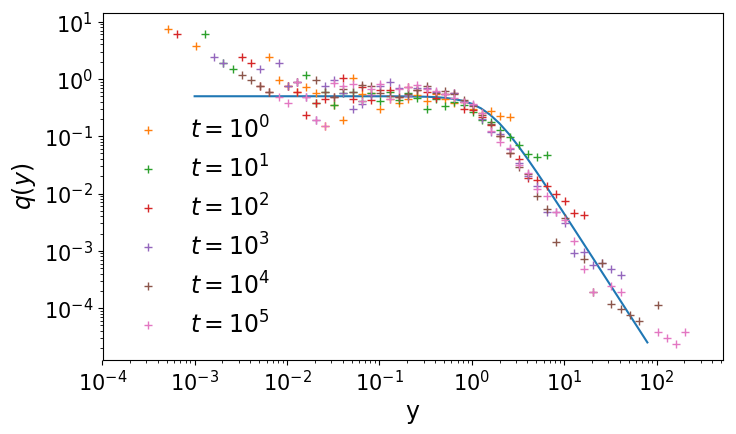}
\caption{Thaler-Dynkin limit Theorem for the Pomeau-Manneville map
(also see in Akimoto and Barkai \cite{AkimotoBarkai})
for $z=3.5$. The rescaled densities of the transformed variable $y$ 
at different times $t$ (points) collapse to a universal curve 
described by Eq. (\ref{eqINF04}) (blue curve).}
\label{fig08}
\end{figure}

When $z>4$ the variable $|v|^2$ is non integrable with respect to the
infinite density. The average of this observable 
is then  obtained using Thaler-Dynkin limit theorem. Dynkin's
result  \cite{Dynkin}  was derived in the context of 
renewal theory in particular the
analysis of the forward recurrence time \cite{GodrecheLuck}, 
while Thaler established the
connection to the underlying transformations \cite{ThalerDyn}. 
We will not go into the
details since they were recently explained in \cite{AkimotoBarkai}. 
Briefly, we define the rescaled  variable  with
$y=|v|(\tilde{a} t)^\alpha /\alpha^\alpha$ and 
the normalized  PDF of $y>0$, in the long time limit,  is
$q(y) =[\sin \pi \alpha/\pi \alpha]/[1 + y^{1/\alpha}]$.
A numerical calculation is shown in Fig. \ref{fig08}.
Averages like $|v|^2$ which is of course proportional  to averages
of  $y^2$ are now
obtained with  this limiting  PDF.
Importantly we have the scaling $|v|\sim t^{-\alpha}$ 
(but this should be used with care, and is limited to the observables 
non integrable with respect 
to the infinite density). It is easy to show that
\begin{equation}
\label{eqINF04}
\langle |v|^2 \rangle  \sim 
\left( {\alpha \over \tilde{a} t} \right)^{2 \alpha}  
\int_0 ^\infty {\sin \pi \alpha \over \pi \alpha} 
{ y^2 \over 1 + y^{1/\alpha} } {\rm d} y .
\end{equation}
So we have $\beta= - 2 \alpha= - 2 /(z-1)$, for $z>4$, as in Eq.
(\ref{eqPM07}). 
Note that the integral in Eq. 
(\ref{eqINF04}) blows up when $\alpha>1/3$, i.e. $z<4$, 
due to the upper limit, 
hence in that case the observable $y^2$ (or $|v|^2$)
is non-integrable with respect to the
Thaler-Dynkin distribution. 
In that sense the infinite density and the Thaler-Dynkin limit theorem
are complimentary to one another, in fact large $y$ behavior of the latter
\begin{figure}[htb]
\includegraphics[width=0.5\textwidth]{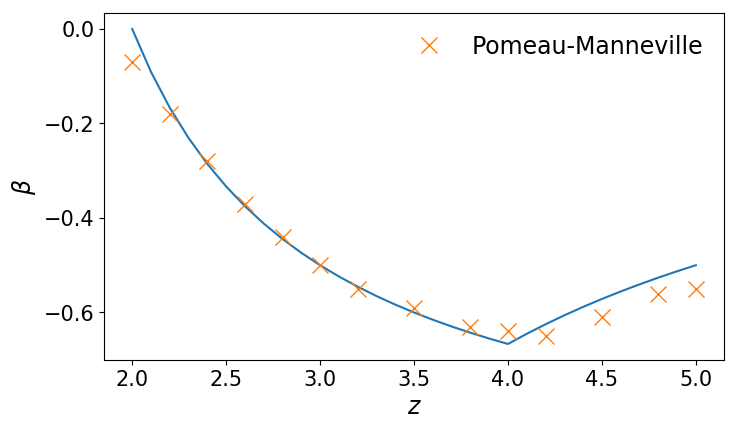}
\caption{The scaling $\beta$ of the velocity displacement 
$\langle v^2\rangle$ was fitted after 
$t=10^6$ iterations of the map. The solid curve shows the analytical 
calculation $\beta$ for $t\rightarrow\infty$ using the infinite 
invariant density ($z<4$) and the Thaler-Dynkin limit theorem ($z>4$)
(see Eq. (\ref{eqINF04})).
The exponent beta is negative since the particle is attracted to the 
unstable fixed points on $v=0$, so $\langle v^2\rangle$ is decreasing with time as 
more particle are accumulated close to the unstable fixed point.}
\label{fig09}
\end{figure}
matches the small argument behavior of the former, 
the existence of two limits
is related to non uniform convergence of the density of $|v|$.
To summarize $\beta$ in our process is given by
\begin{equation}
\label{eqINF05}
\beta=\left\lbrace\begin{array}{ l l l  l}
\nu-2=2\gamma-2  \ & \ \mbox{if \ } \ 2 \gamma> 1 + \alpha \\
\alpha-1  \ & \ \mbox{if \ } \ 2 \gamma<1 +\alpha.  
\end{array}\right. .
\end{equation}
See Fig. \ref{fig09} for the numerical calculation.
Note that the result is identical to the value of 
$\beta$ in Eq. (\ref{eq29t}). We can also see the exponent $\beta$
in the velocity correlation function (see Fig. \ref{fig10}). However, 
we have to be careful about how we make the transition 
$\tau /t\rightarrow 0$.
\begin{figure}[htb]
\includegraphics[width=0.235\textwidth]{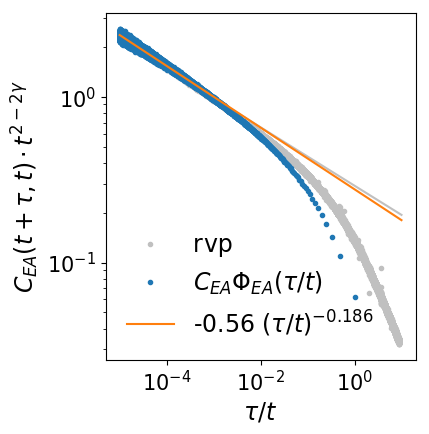}
\includegraphics[width=0.235\textwidth]{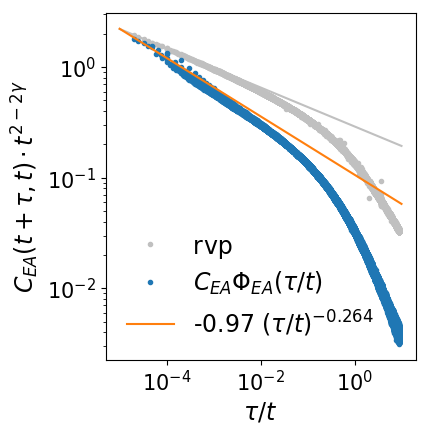}
\caption{Calculation of the exponent $\delta_1$ ($\beta=\delta_1+\nu -2$) 
from the velocity correlation 
function of the Pomeau-Manneville map compared to the data for
the renewal velocity process in Fig. \ref{fig03}.
Left: Scaling function for $\tau=const=10$ and $t\rightarrow \infty$
gives the expected result for $\beta$.
Right: using the same data as in Fig. \ref{fig05}. The
exponent is different from the expected $\delta_1$ and the data
for different $\tau$ does not collapse close
to zero to one curve.
}
\label{fig10}
\end{figure}

\subsection{Scaling of the time average}
Now we have transport  exponents.
We insert Eq. (\ref{eqINF05}) and (\ref{eqPM06}) in Eq. (\ref{eq17}) 
and get for the TA MSD
\begin{equation}
\label{eqPM08}
\langle \overline{\delta^2} \rangle \sim \left\{
\begin{array}{l l}
t^{ {2 - z \over z-1} } \Delta^{ 3 \left( { z-2 \over z-1} \right) } 
& \ 5/2 < z< 4 \\
t^{ - {2 \over z -1} } \Delta^2 & z > 4
\end{array}
\right. .
\end{equation}
\begin{figure}[htb]
\includegraphics[width=0.235\textwidth]{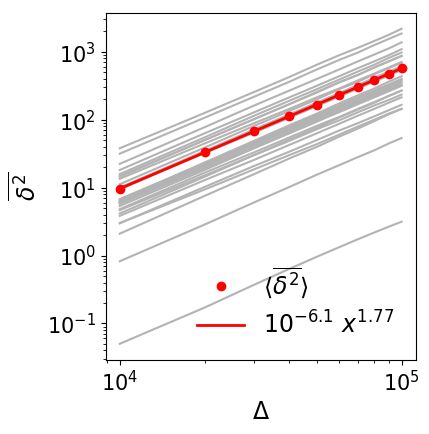}
\includegraphics[width=0.235\textwidth]{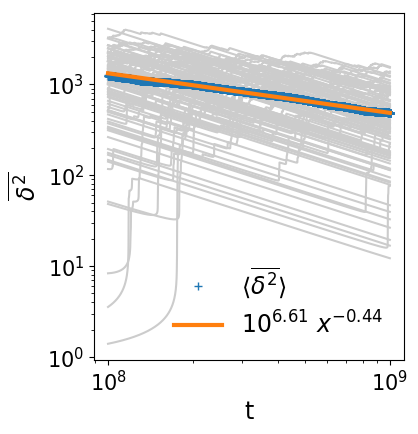}
\caption{TA MSD of the deterministic system: 
Calculation of the dependence on $\Delta$ (left) 
for $z=3.5$, $t=10^{10}$ and $t$ (right) for $z=2.8$, $\Delta=10^6$. 
In both panels we 
present also the EA TA MSD 
(colored lines) which nicely match with  Eq. (\ref{eqPM08}).
We see that the scaling with $\Delta$ is very stable 
within the ensemble (grey curves) and fluctuations are 
only visible in the pre-factor. In contrast for the dependence of 
$\delta^2$ on $t$ the EA is not similar to
one single TA MSD calculation.}
\label{fig11}
\end{figure}
The exponent of $\Delta$ in the TA MSD will now be called $\eta$
i.e. $\delta^2 \propto \Delta^\eta$.
Eq. (\ref{eqPM08}) yields the EA TA MSD. 
What will one observe based on an 
individual trajectory analysis without any ensemble averaging? 
Generating one trajectory at a time using the Pomeau-Manneville map 
we estimate numerically $\overline{\delta^2}$. 
Fig. \ref{fig11} shows that the scaling dependence on 
$\Delta$ is stable even if the pre-factors of the TA MSD fluctuate. 

\begin{figure}[htb]
\includegraphics[width=0.5\textwidth]{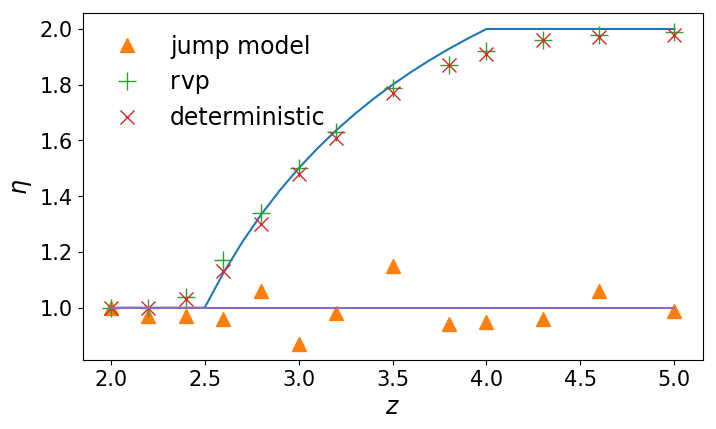}
\caption{Scaling exponent in $\Delta$ of the TA MSD for all three processes}
\label{fig12}
\end{figure}

For $z>2.5$ it is identical to $\nu -\beta$.
If $z$ is smaller than $2.5$ condition (\ref{eq11a}) does not 
hold and thus the scaling Green-Kubo relation is not applicable. 
For $z>4$ the particle exhibits ballistic behavior as the TA MSD 
increases quadratically with $\Delta$ but also aging as 
the $\Delta^2$ pre-factor is shrinking as we increase the observation
time. 
This large $4<z$ limit corresponds to small values of $\alpha$ 
which means  that the particle is getting trapped very close to the
vicinity of the two-sided unstable fixed point on the origin, 
so velocity is effectively decreasing (i.e. $\beta<0$) but still 
the particle remains with $v>0$ or $v<0$ 
without switching sign, for times of the order of measurement time, 
which gives the ballistic like feature of the time averaged MSD.
As we cross to the regime  $2.5<z<4$, $\alpha$ is decreased, the ballistic 
transport turns super-diffusive, but still we have an aging pre-factor.
The results for the scaling are the same that we got for the 
renewal velocity process (see Fig. \ref{fig12}).
The scaling Green-Kubo approach is not valid for $z<2.5$ 
since the condition $\nu-\beta>1$ does not hold there. 
This does not mean that $z<2.5$ exhibits perfectly normal diffusion, 
namely scaling Green-Kubo and standard Green-Kubo relations, 
are not yet the complete story.

\section{Conclusion}

We found a theory that relates the TA MSD to the correlation function 
if the velocity correlation function is scale invariant and the scaling 
exponents $\beta$ and $\nu$ obey Eq (\ref{eq11a}).
The scaling GK relation can be applied for systems where the 
usual Green-Kubo relation does not work.
There are several systems where this approach could be useful, 
like cold atoms diffusivity on 
optical latices, active transport in cells and blinking quantum 
dots \cite{DechantPRX}.

With the scale invariant GK relation we
can calculate the TA MSD
of renewal velocity processes. 
Depending on the waiting time distribution 
and the velocity scaling 
the process might show either subdiffusion, 
normal diffusion, superdiffusion or ballistic motion. 
A transition occurs when $2\gamma=\alpha +1$.
The most simple version of such a renewal velocity process where 
$v_{\gamma,t}=\pm 1$ 
can easily be related to a random walk model where jump length and 
waiting time distribution are the same.
We explored cases where the scaling exponents of the EA MSD 
are identical for the random walk and velocity approach, 
still the exponents of the TA procedure for the two models differ. 
This is somewhat surprising as on the renewal times, 
the path of the two processes is identical. 
So the TA MSD
is sensitive to the choice of the model 
and to the precise definition of the paths (see Fig. \ref{fig04}). 
In the context of Langevin equation with multiplicative noise, 
the difference between Ito and Stratonovich calculus,
which is also related to the precise definition of stochastic paths, 
is well documented \cite{itostra}. In this manuscript, 
within the context of anomalous diffusion, the  exact
shape of the path is also extremely important
(unlike normal processes).
The effect stems from the fact that in the 
measurement time interval (0,t) we have a single flight or waiting 
event that dominates the trajectory, 
in the sense that these are of the order of measurement time t \cite{GL1}.
Hence, the precise definition of the path of the particle in this interval,
crucially influences the output of the time averaged procedure, 
but not the ensemble average, since the latter is a measure of 
where is the particle at the moment of observation, 
while the former a functional of the whole path. 

We also investigated a deterministic model of a diffusion process 
generated by a symmetric version of the Pomeau-Manneville map.
Here numerical calculations of the correlation function
showed that the exponent $\delta_1$ depends on how the limit
$\tau /t\rightarrow 0$ is approached. 
Analytical calculations 
can be done using infinite ergodic theory. 
The result for the scaling exponents is the same 
as for the stochastic velocity model.
In the range where the integration of $v^2$ with 
respect to the infinite invariant density diverges, the 
Thaler-Dynkin law can be applied.

\section*{Acknowledgments}
This work was supported by the Israel science foundation.

\end{document}